\documentclass{aa}
\usepackage{natbib}
\bibpunct{(}{)}{;}{a}{}{,} 
\usepackage{algorithm}
\usepackage[noend]{algpseudocode}
\usepackage{graphicx}
\usepackage{float}
\graphicspath{{img/}{../img/}}

\usepackage{caption}
\usepackage{subcaption}

\usepackage{txfonts}
\usepackage{hyperref}
\hypersetup{
	colorlinks=true,
	linkcolor=red,
	urlcolor=blue,
	citecolor=blue
}

\begin{document} 

\title{ExoplANNET: A deep learning algorithm to detect and identify planetary signals in radial velocity data}
\titlerunning{ExoplANNET: deep learning for exoplanet detection in RV data}
\author{
	L.~A.~Nieto\inst{1, 2} \&
	R.~F.~D\'iaz\inst{2} 
}
\institute{
Gerencia de Tecnología de la información y de las Comunicaciones (GTIC), Subgerencia Vinculación y Desarrollo de Nuevas Tecnologías de la Información, DCAP-CNEA. Centro Atómico Constituyentes, Av. Gral. Paz 1499, (1650) Buenos Aires, Argentina.
\and
International Center for Advanced Studies (ICAS) and ICIFI (CONICET), ECyT-UNSAM, Campus Miguelete, 25 de Mayo y Francia, (1650) Buenos Aires, Argentina.}
\date{Received TBC; accepted TBC}
\offprints{L.A Nieto (lnieto@unsam.edu.ar)}



 \abstract{

 	{The detection of exoplanets with the radial velocity method consists in detecting variations of the stellar velocity caused by an unseen  sub-stellar companion. Instrumental errors, irregular time sampling, and different noise sources originating in the intrinsic variability of the star can hinder the interpretation of the data, and even lead to spurious detections. }
 		
	{In recent times, work began to emerge in the field of extrasolar planets that use Machine Learning algorithms, some with results that exceed those obtained with the traditional techniques in the field. We seek to explore the scope of the neural networks in the radial velocity method, in particular for exoplanet detection in the presence of correlated noise of stellar origin.}

	{In this work, a neural network is proposed to replace the computation of the significance of the signal detected with the radial velocity method and to classify it as of planetary origin or not. The algorithm is trained using synthetic data of systems with and without planetary companions. We injected realistic correlated noise in the simulations, based on previous studies of the behaviour of stellar activity. The performance of the network is compared to the traditional method based on null hypothesis significance testing.}
 	
	{The network achieves 28 \% fewer false positives. The improvement is observed mainly in the detection of small-amplitude signals associated with low-mass planets. In addition, its execution time is five orders of magnitude faster than the traditional method.}

  {The superior performance exhibited by the algorithm has only been tested on simulated radial velocity data so far. Although in principle it should be straightforward to adapt it for use in real time series, its performance has to be tested thoroughly. Future work should permit evaluating its potential for adoption as a valuable tool for exoplanet detection.}
 }

   \keywords{
   	Methods: data analysis --
   	Techniques: radial velocities --
   	Planets and satellites: detection
    }

   \maketitle
%
\section{Introduction}
The study of extrasolar planets is a relatively new field of research. Although the first evidence of the existence of this type of bodies dates from 1917 \citep{Landau:2017}, it was not until the 1990s that the first confirmed detections took place. In 1992, by analysing the variations in the period of the pulses received from the radio millisecond pulsar PSR1257+12  \citet{Wolszczan1992} concluded that at least two Earth-mass planets are in orbit around the pulsar. Three years later, \citet{1995Natur.378..355M} discovered the first exoplanet orbiting a solar-type star, 51 Peg b, by measuring the variations in the line-of-sight (radial) velocity of the host star induced by the unseen companion. In the years that followed this first detection, the radial velocity technique allowed unveiling a large number of planet candidates and information-rich systems, some with masses as small as a few times the mass of the Earth \citep[e.g.][]{lovis2006, mayor2009, wright2016, angladaescude2016a, astudillodefru2017c, astudillodefru2017b, feng2017,delisle2018, bonfils2018, diaz2019, zechmeister2019, dreizler2020} and/or that are promising candidates for future atmospheric characterisation \citep[e.g.][]{bonfils2018, diaz2019}.

These discoveries were in large fraction possible thanks to the continuous improvement in instrumentation that provides an ever-increasing precision in the measurement of the radial velocity of the stars. Going from the pioneer ELODIE spectrograph \citep{baranne1996}, to HARPS \citep{mayor2003}, SOPHIE \citep{perruchot2008, bouchy2013}, and CARMENES \citep{quirrenbach2014}, and finally, the ultra-high precision spectrographs such as ESPRESSO \citep{pepe2021} and EXPRES \citep{jurgenson2016, blackman2020}, the precision was improved by over two orders of magnitude, attaining now the level of 10 cm s$^{-1}$.

However, planet detection with these observations are not limited only by instrumental precision, but by intrinsic variability in the star. Even for the least active, slowly-rotating stars, the phenomena collectively called \emph{stellar activity} can produce spurious radial velocity variations with amplitudes of up to a few meters per second, and timescales ranging from a few minutes \citep[pulsations and granulation; e.g.][]{dumusque2011a} to decades \citep[activity cycles; e.g.][]{lovis2011b, diaz2016a}. Particularly worrying are the variations produced by the rotational modulation of the star, as they tend to exhibit power in the same frequency range as some of the most interesting planetary candidates \citep[e.g.][]{Saar_1997, boisse2009, dumusque2011b, nielsen2013}. The main consequence of the influence of stellar activity is the difficulty to detect exoplanets producing RV variations smaller than 1-ms$^{-1}$.

Some of the most relevant stellar phenomena that can generate this type of noise are:
	\begin{enumerate}
		\item{\textit{Stellar oscillation}:} Pressure waves (p-modes) propagate at the surface of solar-type stars causing the contraction and expansion of the outer layers over timescales of a few minutes (5-15 min for the Sun, \cite{Broomhall2009,schrijver_zwaan_2000}). The radial-velocity signature of these modes typically varies between 10 and 400 cm s$^{-1}$, depending on the star type and evolutionary stage \citep{schrijver_zwaan_2000}.

		\item{\textit{Granulation}:} Various convective motions in the photosphere cause this phenomenon. These \textit{granules}, that emerge from the interior of the star into the photosphere, are hotter than those that cool down and descend. This produces a spurious redshift \citep{refId0, dumusque2011a} that changes depending on the convection pattern and can range from a few minutes to about 48 hours.
		
		\item{\textit{Rotational modulation}:} The rotation of the star can transport various structures on the surface, causing them to appear and disappear at regular intervals breaking the flux balance between the red-shifted and the blue-shifted halves of the star and creating the illusion of a stellar wobble \citep{Saar_1997, Lagrange2010, dumusque2011b}.
		
		\end{enumerate}

From a statistical standpoint, one of the main issues is that these phenomena produce correlated error terms, that invalidate some of the most commonly used techniques for planet detection that rely on ordinary least squares, such as the standard periodogram analyses \citep{baluev-fap, zechmeisterkurster2009}. This problem has often been approached by including correlated noise errors in the modelling of the data. In particular, the use of Gaussian process regression \citep{rasmussenwilliams2005} has been widely adopted by the community \citep[e.g.][just to cite a few among a vast body of literature using this technique]{haywood2014, rajpaul2015, yu2017, cloutier2017, persson2018, bonfils2018, diaz2019, luque2019, suarezmascareno2020}, fuelled by the availability of specific computer code to perform the necessary calculations effectively \citep{foreman-mackey2017, espinoza2019, delisle2022}. However, methods to compare models including this kind of error terms can be time-consuming or unreliable \citep{nelson2020}.

A different approach is to bypass the problem of explicitly modelling the effects of activity together and rely instead on machine learning models to perform the detection and classification tasks. This has been used mostly for the detection and veto of transiting planet candidates with photometric time series \citep[e.g.][]{zucker2018}, mainly from the space missions \emph{Kepler} \citep[e.g.][]{armstrong2017, Shallue_2018, pearson2018, ansdell2018}, \emph{K2} \citep[e.g.][]{dattilo2019}, or \emph{TESS} \citep[e.g.][]{yu2019, osborn2020, rao2021}, but also from ground-based transit surveys such as the Next Generation Transit Survey \citep[e.g.][]{McCauliff_2015, armstrong2018}. These models mostly rely on deep convolutional networks \citep[e.g.][]{lecun1998}, that have proven extremely proficient in many fields, mainly computer vision \citep[e.g.][]{krizhevsky2012, he2016} and natural language processing \citep[e.g.][]{peters2018, devlin2019}. It is not a surprise that these methods have been advanced mostly for spaced-based photometric surveys, as these missions provide rich datasets that are needed to train this kind of machine learning models. But machine learning has also been employed in the context of exoplanets to study their atmospheres \citep[e.g.][]{Marquez-Neila, 2016ApJ...820..107W}, or to classify planets according to their potential habitability \citep[e.g.][]{Basak}. Recently, \citet{debeurs2020identifying} used neural networks to remove the stellar activity signal from simulated and real RV observations and showed how this can help in exoplanet detection.

In this article, we train a convolutional neural network (CNN) with simulated data to perform the detection of extrasolar planets in radial velocity time series. We frame the question as a classification problem and train the model to distinguish periodograms with and without planetary signals. We show that this method produces better results than the traditional periodogram analyses and correctly identifies more low-mass planets, being also much faster.
\section{Planet detection with GLS periodograms}
The radial velocity method consists of detecting variations in the line-of-sight projection of the star's velocity, induced by the presence of an unseen companion. 

\subsection{Periodograms}
\label{sec:periodograms}

Periodograms are often employed to find periodic signals in unevenly sampled time series as those typically obtained in RV surveys. The Lomb-Scargle periodogram \citep{periodogram, review_signal} has been used extensively to detect periodic signals in RV datasets. In its simplest form, it works in a Fourier-like fashion: the user defines a series of candidate frequencies $\{\omega_i\}$, with $i=1,...,N_f$, and the method fits the radial velocity data, $v(t)$, using a sine-cosine base at the candidate frequency. The better the model at this frequency fits the data (compared to a constant model without variability), the higher the power attributed to the candidate frequency \citep[for details, see][]{lomb1976, scargle1982, baluev-fap}. Indeed, the power at frequency $\omega_i$ is 

\begin{equation}
P(w_i) = \frac{\chi^2_0 - \chi^2_i}{\chi^2_0}\;\;,
\label{eq.power}
\end{equation}
where $\chi^2_i$ is the chi-square statistics for the model at frequency $w_i$. If we call $f_i(t)$ the model prediction at time $t$, 

$$
\chi^2_i = \sum_{j=1}^N \frac{\left(v(t_j) - f_i(t_j)\right)^2}{\sigma_j^2}\;\;,$$
where the sum runs over the $N$ data points in the time series.

\begin{figure*}
     \centering
     \begin{subfigure}[b]{0.48\textwidth}
         \centering
 		    \includegraphics[width=1\linewidth]{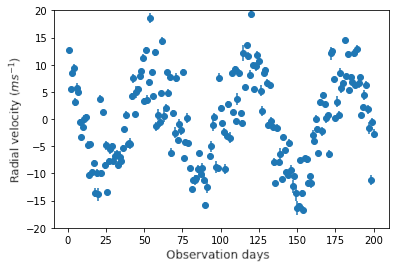}\hfill%
         \caption{Radial velocity measurements.}
         \label{fig:_1}
     \end{subfigure}
     \hfill
     \begin{subfigure}[b]{0.48\textwidth}
         \centering
			\includegraphics[width=1\linewidth]{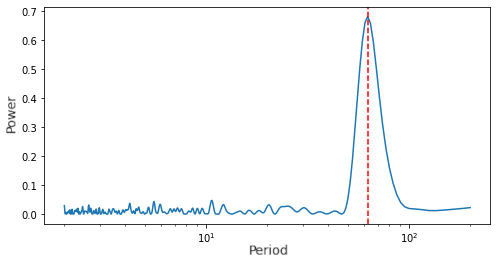}
         \caption{Periodogram. The red line marks the peak that represents the period with the highest power}
         \label{fig:_2}
     \end{subfigure}
     \hfill
        \caption{Radial velocity measurements and their periodogram.}
        \label{fig:rv_pg}
\end{figure*}

A vector is generated with the power of each candidate frequency. The location of the highest power indicates the frequency of a potential periodicity present in the data. Figure \ref{fig:rv_pg} shows a time series (with a clear periodic signal) and the resulting periodogram. Note that the width of the peaks expressed in frequency depends solely on the time span of the observations $\Delta T$, and can be expressed 
$$
\Delta_\omega = 1/\Delta t\;\;.
$$

\subsection{False alarm probability (FAP)}

Once the periodogram has been constructed, and a candidate period is identified, a tool is needed to help decide whether the observed peak is statistically significant to be considered a periodic signal. This question is traditionally framed as a null-hypothesis significance test, in which the null hypothesis ($H_0$, "there is no planet") is tested. More precisely, in the analysis of the Lomb-Scargle periodograms, the null hypothesis is often that the observations come from uncorrelated Gaussian noise (i.e., ``there is no planet''). The power of the largest peak is used as the test statistics, for which a $p$-value is computed\footnote{In the exoplanet literature, this is often referred to as False Alarm Probability (FAP). Although we deem this terminology misleading we will use $p$-value and FAP interchangeably.}. If the $p$-value is small enough, then the null hypothesis is rejected. The critical value below which a $p$-value is considered as evidence for rejecting the null hypothesis is usually between $10^{-3}$ and 0.1. In many planet search programmes, a FAP value of 0.01 is often used as a threshold below which detected peaks are considered significant \citep[e.g.][]{howard2010, bonfils2013}. Based on these previous studies, the critical value --which we call $FAP_*$ --will be taken as 0.05 with $\log_{10} (FAP_*)$ rounded to -1.3.

To compute the $p$-value (FAP) one can take an analytical or numerical approach, using a Monte Carlo algorithm. The analytical expressions \citep[e.g.][]{scargle1982, baluev-fap, VanderPlas_2018} have the advantage of being much faster to compute than Monte Carlo approaches. However, these expressions rely on assumptions and approximations that are only reliable in certain regimes. For example, they often rely on the assumption that the errors are uncorrelated and normally distributed. Following the comments in \citet{baluev-fap} we considered that, if the $p$-value is greater than or equal to $log_{10}$ (0.01), the analytical expression is no longer reliable and the other method should be preferred. On the other hand, the Monte Carlo approach samples the data points keeping the observing times and is therefore probably more reliable even for situations where the assumptions on which analytical expressions are based do not hold. In this work, a number of noise realizations are sampled to compute the $p$-value of the largest peak power with this approach. As we work with simulations, the correct noise distribution is used. This is the most favourable situation which is rarely, if ever, given with real datasets. The Monte Carlo approach requires recomputing hundreds or thousands of periodograms, which makes it a much slower method than the analytical approach, especially when taking into account that we want to analyse a large amount of time series.

As in any null-hypothesis significance test, we are exposed to type I and II errors \citep[e.g.][]{frodesen}. Type I, or ``false alarm errors'', occur when $ H_0 $ is incorrectly rejected (that is, we decided a peak is significant when it is not). Type II errors are when we fail to reject $ H_0 $ when in fact a planet is present. These errors are also often referred to as false positives and false negatives. 

When using this approach, one still needs to decide whether the detected signal is of planetary origin or caused by activity or systematic effects. Often, time series of activity indexes are employed to verify that the candidate frequency is not present in these series, which is regarded as evidence for the planetary interpretation of the signal. However, there is not a universally accepted way to decide whether a signal is produced by a planet.

\section{Simulations}

To train, validate and test the network a data set of sufficient size is required. In principle, it is not known how large this data set needs to be, but the larger the set with which the network is trained and validated, the better the results it is expected to provide (see \citeauthor{banko2001} \citeyear{banko2001} for a famous example, and the discussion by \citeauthor{halevy2009} \citeyear{halevy2009}). 
 
The data set not only needs to be large but also representative, in terms of number of points, variability, presence of periodic signals, etc., of the cases that the network will encounter when used in ``production stage'', i.e. when used to classify actual time series. The available radial velocity time series issued from large scale surveys are too few and too diverse for an efficient training. They do not have enough representatives of each type of variability and planetary system architecture. Besides, not knowing precisely the number of planets in the system means we would also have to deal with the problem of inaccurate labelling, i.e. not being able to precisely convey to the algorithm the class to which a given data set belongs.

The alternative is to train and evaluate the network on synthetic RV time series, and later study its performance on real data sets. If the simulated data are realistic enough, this may be enough, but most probably, an adaptation of the network architecture would be needed. Artificial data have many obvious advantages: being able to produce a large number of time series, and precisely knowing the number of periodic signals in them is fundamental. On the other hand, one clear disadvantage is that the results from our study, at least in principle, will be as relevant as our simulations are realistic.

Our network takes periodograms as inputs. We proceed by first generating synthetic time series, from which the Generalised Least-Squared \citep[GLS][]{zechmeisterkurster2009} periodogram is produced. The reason for this is that while we have relatively good models to simulate time series, the problem of directly simulating periodograms is much tougher. The steps of the simulation process are detailed in the subsections below.

\subsection{Time array}
The synthetic time series are sampled using 200 quasi-uniformly distributed points. The times are defined by adding random variables drawn from a zero-centred normal distribution with width 0.1 to an evenly spaced time array. Although not completely realistic, this sampling roughly represents an intensive observing season with nightly observations on each target\footnote{The sampling may be seen as done in sidereal days}.

\subsection{Intrinsic errors}

Intrinsic error terms\footnote{Here we call error terms to the unknown term that makes the observed velocities deviate from the model. In contrast, the term uncertainties is reserved to the experimental error associated with each measurement.} are simulated by randomly drawing samples from an uncorrelated multivariate zero-centred normal distribution, with variances informed by the statistics of the uncertainties of the HARPS high-precision survey \citep[e.g.]{harps, Diaz2016TheHS}. For each simulated star, a star is randomly drawn from the subset of HARPS targets that were observed more than forty times by the survey. Using the observations of this star, the mean velocity uncertainty, $\bar{\sigma_v}$, is computed. Then, 200 samples are drawn from a normal distribution centred on $\bar{\sigma_v}$ and with a width of 0.30  ms$^{-1}$, which is a typical value for the dispersion in the velocity uncertainties of observed stars. Negative values are replaced by $0.5 \bar{\sigma_v}$. In this way, each data point has a slightly different uncertainty and error term, as is usually the case in real time series.

\subsection{Activity signals}

It has already been mentioned that there are several components of stellar noise that can alter radial velocity measurements. Simulated effects of pulsations and granulations were based on the work by \citet{dumusque2011a}. The authors studied the power spectrum of five well-observed solar stars, the spectrum is fitted with a Lorentzian component that represents the pulsations and three components for the granulation, mesogranulation and supergranulation. Using the fitted parameters defined in there, the power spectrum is constructed as the sum of these four components, with randomly chosen phases for each component. The resulting power spectrum is used to generate the synthetic radial velocity measurements.

The effect of the rotational modulation was simulated using samples from a Gaussian process with a covariance function generated with the pseudo periodic kernel:

$$
k_\mathrm{QP}(t_i, t_j) = A^2 \exp\left(-\frac{(t_i - t_j)^2}{2\tau^2} - \frac{2}{\epsilon}\sin^2{\left(\frac{\pi (t_i - t_j)}{\mathcal{P}}\right)}\right)\
$$

This \textit{kernel} is used widely in the exoplanet literature to model this kind of effect\footnote{See references in the Introduction for a number of examples of articles using Gaussian Processes with this kernel.} and has been used to infer the rotational period and other parameters of astrophysical interest related to stellar activity and rotation \citep[e.g.][]{giles2017, angus2018} This \textit{kernel} has 4 hyperparameters: $A$, the  covariance amplitude; $\mathcal{P}$, the recurrence time, which would be given by the rotation period; $\tau$, the decay time, associated with the average lifetime of an active region; $\epsilon$, the structure factor, a parameter associated with the number of active regions that occur simultaneously.

To define the values of these hyperparameters for each simulated time series we resorted again to the HARPS survey. All stars in the survey have an estimate of their rotation period obtained from a measurement of their magnetic activity using the calibration by \citet{Mamajek_2008}. For each simulated velocity series, the value of $\mathcal{P}$ was randomly chosen from the sample of estimated rotational periods for HARPS stars. The remaining parameter values were sampled from distributions:
\begin{itemize}
	\item[]{$A \sim$} gamma distribution, $\Gamma$(2.0 , 0.5). 
	\item[]{$\tau\sim$} normal distribution, $\mathcal{N}(3*\mathcal{P}$ , 0.1*$\mathcal{P})$.
	\item[]{$\epsilon\sim$}  uniform distribution, $\mathcal{U}$(0.5 , 1.0).
\end{itemize}

Adding the intrinsic errors to the activity signals makes the time series of radial velocities without a planet $rv_{wp}$.

\subsection{Planets}

The contribution of planets to the velocity time series is generated assuming circular orbits. Additionally, we disregard mutual interaction in multi-planetary systems.

Therefore, we need three parameters: the orbital period, $P_{pl}$, the semi-amplitude of the variation $K$, and the time in which the variation is zero, $T_0$. The value of $P_{pl}$ was chosen randomly between $10 * \delta$ and $\Delta_t/2$, where $ \delta$ is the minimum distance between two points in the time series, and $ \Delta_t$ is the total duration of the series. The value of $K$ was drawn from a log-uniform distribution between 0.1ms$^{-1}$ and 10ms$^{-1}$. Finally,  $T_0$ was randomly chosen as a moment within an orbital period.

\begin{figure*}
	\centering
	\begin{subfigure}{.33\textwidth}
		\centering
		\includegraphics[width=1\linewidth]{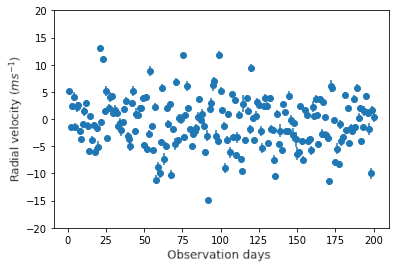}
		\caption{Stellar noise.}
		\label{fig:noise_wo_pl}
	\end{subfigure}%
	\begin{subfigure}{.33\textwidth}
		\centering
		\includegraphics[width=1\linewidth]{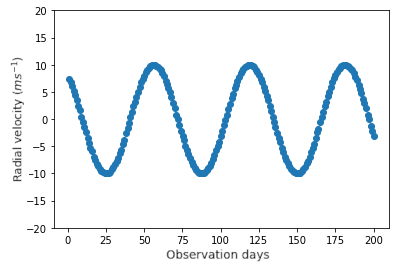}
		\caption{Planetary signal.}
		\label{fig:rv_pl}
	\end{subfigure}
	\begin{subfigure}{.33\textwidth}
		\centering
		\includegraphics[width=1\linewidth]{vel_rad_period_0}
		\caption{Composition.}
		\label{fig:vel_rad_compuesto}
	\end{subfigure}
	\caption{Composition of stellar noise with a planetary signal with a period of 62 days and an amplitude of 10ms$^{-1}$.}
	\label{fig:vel_rad_proceso}
\end{figure*}

The variation then takes the form:
$$ rv_{pl}  = K \sin \left( \frac{ 2 \pi (t - T_0) }{ P_{pl} } \right) $$
Once the vector $rv_{pl} $ has been obtained, it was be added directly to $rv_{wp}$ to get the time series with planet (see figure \ref{fig:vel_rad_proceso}).

In line with our decision to neglect mutual interactions, to model multi-planetary systems, the procedure is repeated for each planet and the contributions were added.

\subsection{Periodograms}

GLS periodograms were computed for each simulated time series on a grid of uniformly-sampled orbital frequencies, between $\delta_\nu$ and 0.5 d$^{-1}$, with a constant step of $\delta_\nu$/10 (i.e. using an oversampling factor of ten), where $\delta_\nu$ is the inverse of the duration of the time series:

$$\delta_\nu = \frac{1}{\max(t) - \min(t)}  $$

\subsection{Datasets}
To train, validate and test the neural network, several datasets are needed to guarantee the robustness of the proposed solution and the quality of the results.

The usual methodology when working with periodograms is to analyze the maximum peak and define if it comes from a significant signal. If it is the case, the signal is removed from the original time series, and a new periodogram is generated using the model residuals, where a new maximum peak is identified and analysed in a similar fashion. This means that the network will not only receive the original periodogram as input, but also subsequent periodograms obtained in this manner. So, in addition to simulating periodograms of stars with up to four planets, we also constructed additional periodograms by removing the signal at the largest peak frequency (regardless of whether it was significant or not). This process was performed four times for each original signal. This not only serves to have more varied periodograms but also increases the size of the training data.

This particular type of problem where an input is considered and has to be assigned to one of two classes (here, planet or not planet) is called a binary classification problem. In this type of problem, it is very important to consider the balance of the classes in the data set --i.e. the fraction of instances from each case-- and the metrics used during training. For example, if an algorithm is trained with many more cases of a given class, it may be biased against predicting the minority class. This may lead to a model with a relatively good accuracy (i.e. the fraction of cases correctly classified), but ultimately useless.

This was considered when constructing the training, validation, and test sets. The first two were for training and parameter adjustment, and the final one was for testing the complete method. For all the sets the same number of stars with 0, 1, 2, 3, and 4 planets were generated and all their periodograms, constructed as described above, were used for parameter tuning and testing. Due to the nature of the generation process, these sets are naturally unbalanced toward negative cases. Consider that from each system, four periodograms are generated independently of the number of planetary signals injected. Then, for each star without planets, for example, four periodograms with negative labels will be produced. In addition, for systems with planets, it may occur that the planetary frequency is not at the maximum peak, which would lead to another negative label, even in the presence of a planet in the simulation. This was not considered a problem for data sets used for evaluation because they are seen as a realistic application. For the training set, however, a balanced data set is more important, periodograms were randomly selected to have near half-positive and half-negative cases. 

In summary, three data sets were generated. In all cases, we simulated an equal number of stars with 0, 1, 2, 3, and 4 planets.
\begin{itemize}
	\item{\textit{Set 1 - 3425 stars}:} from which 13700 periodograms were generated with the process described above. This set was balanced by randomly selecting positive and negative cases until 40\% of positive cases was reached. This set was used for train and validation (see Sect.\ref{sec:train}). 
	\item{\textit{Set 2 - 2550 stars}:} from which we get 10000 periodograms. This set was used for comparisons between methods and threshold search (see Sects.~\ref{lbl:rendimiento} and \ref{lbl:thrsearch}).
	\item{\textit{Set 3 - 5000 stars}:} which lead to 20000 periodograms. It was used to apply the \textit{Virtual Astronomer} to each star and to analyze the distribution of output values for the FAP and the CNN (see Sects.~\ref{lbl:performanceontest} and \ref{lbl:astro_virtual}).
\end{itemize}	
\section{Convolutional Neural Network \label{sec:train}} 

The neural network (nicknamed \texttt{ExoplANNET}) was implemented using the Keras package for \texttt{python}, with Tensorflow as backend. It receives one GLS periodogram, represented in a one-dimensional array of data, together with two additional characteristics: the power and position of the largest peak in the 1D array. The input GLS periodogram was sampled in a grid of 990 frequencies. The frequency grid was fixed and was the same for all simulated velocity datasets. The output is a single value representing the probability that the data was originally created from a periodic signal (in addition to the activity signals). 

A natural solution in this setting is to use convolutional networks in one dimension.\footnote{In section \ref{apendix} there is a brief explanation of this type of network and its use in data that have a spatial relationship, such as images or periodograms.} The architecture is quite standard. The hidden layers will be composed of convolutional layers, interleaved with max pooling layers. Finally, a series of dense layers are implemented, interspersed in turn with dropout layers. All layers use ReLU activation functions except the output neuron which uses a sigmoid activation function for classification that returns a real value between 0 and 1.

The network is trained using the cross-entropy loss function,
$$
E = \sum_{i=1}^N \left[t_i \log(p_i) + (1 - t_i) \log(1 - p_i)\right]\;\;,
$$
where the sum is over all the instances (of a batch) of the dataset, $p_i$ is the predicted probability of belonging to the positive class, and $t_i$ is the target variable (1 if the data contains a periodic signal; 0 otherwise). In each training epoch, the dataset is divided into batches of 128 periodograms, and the weights (parameters) of the model are adjusted using the Adam optimizer, with gradients computed based on the partial loss function. This is standard procedure in machine learning literature and has been shown to increase the performance and convergence of the optimization process \citep[see, for example,][]{bishop2007}.

The evolution of the training was monitored using the \textit{F-measure} (or F-score), which is the harmonic mean of precision ($P$) and recall ($R$), $F_1 = 2 (P R)/(P + R)$. In turn, precision and recall are defined:

\begin{itemize}
	\item{\textit{Precision}}: is the ratio of \textit{true} positive cases among those that were \textit{marked} as positive. In our case it would be:
	$$ P = \frac{\text{\{Real planets\}}\cap \text{\{Marked as planet\}}} {\text{\{Marked as planet\}}}$$

	\item{\textit{Recall}:} It is the ratio of \textit{true} positive cases detected among the positive \textit{existing}. 
	$$ R = \frac{\text{\{Real planets\}} \cap \text{\{Marked as planet\}}} {\text {\{Real planets\}}}$$
\end{itemize}
Unlike the \textit{accuracy}, the F-score is appropriate even in unbalanced cases.

\begin{figure}
	\centering
	\includegraphics[width=0.55\linewidth]{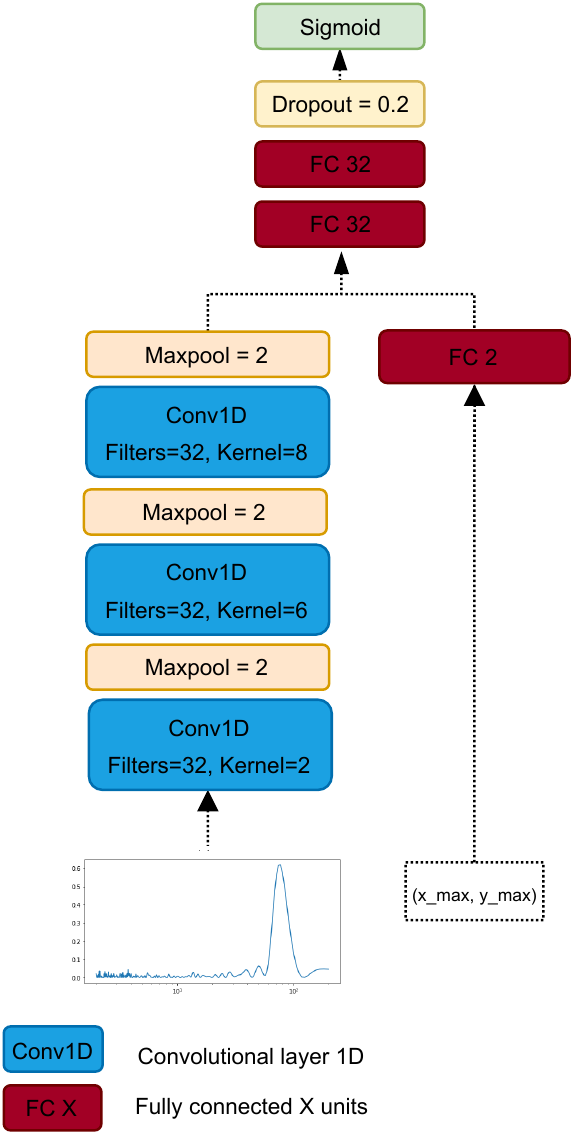}
	\caption{Architecture of the ExoplANNET neural network. Blue cells represent convolutional layers; pale orange cells are MaxPooling layers and dark red cells are fully connected layers. In addition, we show the single DropOut layer in yellow and the output layer in green.}
	\label{fig:network}
\end{figure}

Different network configurations were tested and evaluated with the validation set. The final architecture achieved an F-score of 0.88 in the validation set. A diagram of retained architecture is shown in figure \ref{fig:network}. The trained network is provided in h5 format in a GitHub repository\footnote{https://github.com/nicklessagus/ExoplANNET}, together with a sample of labelled periodograms for testing.

\section{Results}

\subsection{Predictive performance}
\label{lbl:rendimiento}

For a given periodogram, both the classical FAP computation and our CNN output a number.  In the former, it is the log probability that a peak with at least as much power as the largest peak observed in the periodogram appears under the null hypothesis (i.e. only noise --albeit correlated-- is present in the time series). In the latter, it is a value between 0 and 1 that can be interpreted as the probability that the periodogram contains a \emph{bona fide} planetary signal. Both implementations require the definition of a threshold value if they are to be used for taking decisions, such as defining whether a planet is present in the data or not. However, note that the FAP method is only intended to assess the significance of a signal, while the neural network also aims at providing a classification of the nature of the signal.

Before setting this threshold, an analysis of the overall effectiveness of each method can be made and compared with each other, using \textit{precision-recall} curves. From the definitions seen in the previous section, it is trivial to see that both metrics vary between 0 and 1, the intuition behind precision is that it seeks to minimize false positives and recall does so with false negatives.

\begin{figure}[H]
	\centering
	\resizebox{\hsize}{!}{\includegraphics[width=0.7\linewidth]{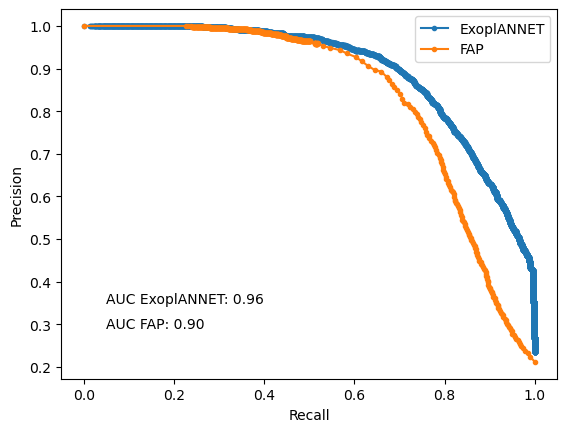}}
	\caption{Precision-Recall Curve and AUC of both methods over the periodograms of \textit{set 2}.}
	\label{fig:pr-curve}
\end{figure}

The precision-recall (PR) curve (Fig. \ref{fig:pr-curve}) shows the tradeoff between these metrics. These curves are constructed by varying the threshold for detection and computing the metrics at each value. Ideally, the curves should pass as close as possible to the upper right edge of the plot, which represents a system with perfect recall and no false positives. Therefore, the greater the area under this curve, the better the overall performance of the algorithm. The area under the curve of the PR curve (PR-AUC) is useful to compare two different solutions to the same problem.

In Figure \ref{fig:pr-curve} we present the PR curve and PR-AUC metric for the classical FAP method over the periodograms of \textit{set 2}. We see that the network performs systematically better than the traditional method. The AUC metrics are 0.90 and 0.96 for the FAP and \texttt{ExoplANNET} methods, respectively.

\begin{figure*}
	\includegraphics[width=1\linewidth]{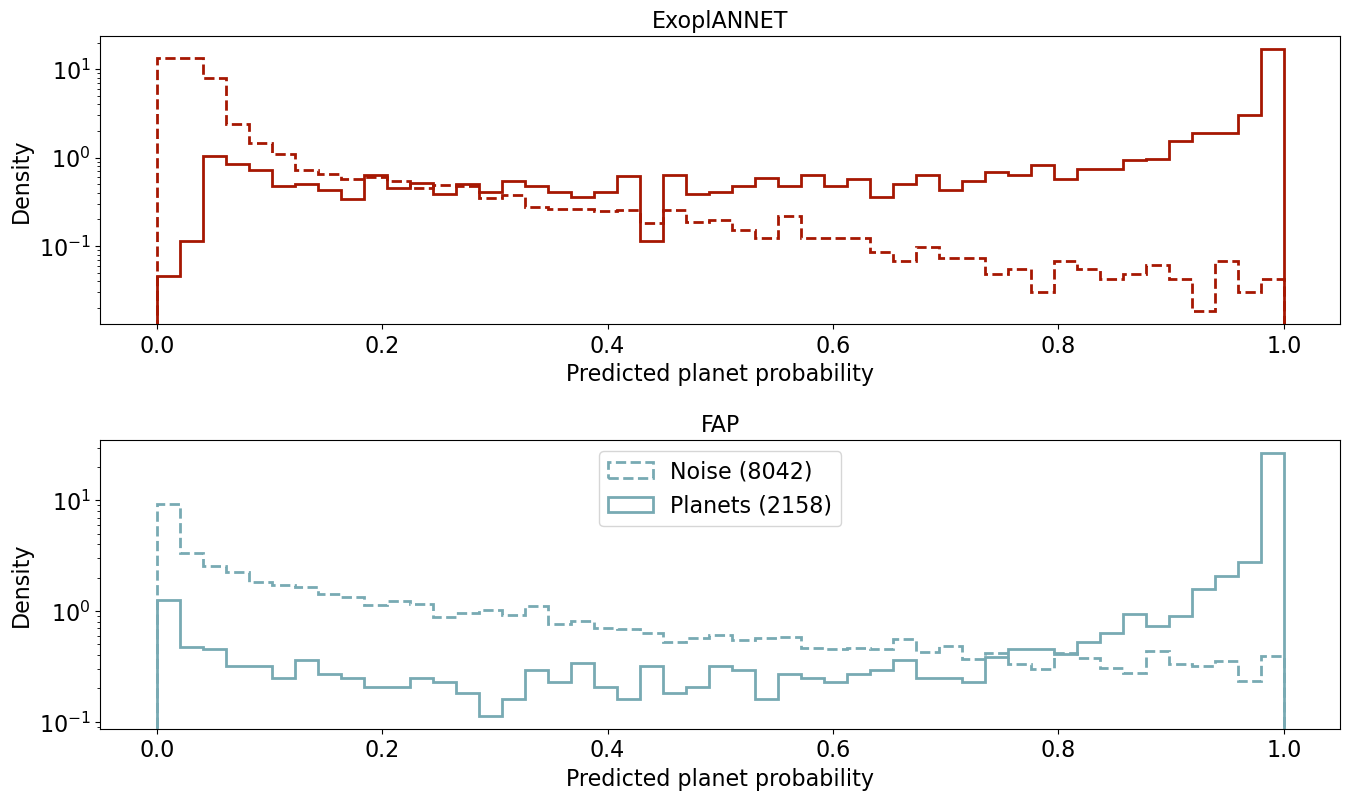}
	\caption{Distribution of output values over the periodograms of set 2 (probabilities and FAP values for the CNN and classical analysis, respectively). For each algorithm the distribution of data labelled as planets and as noise are shown with different line styles. As there are 15587 periodograms without planets for the 4413 planetary periodograms, the distributions are normalised for easier comparison.}
	\label{fig:casosxprob}
\end{figure*}

In  Figure \ref{fig:casosxprob} we present the distribution of the output values (i.e. FAP values and planet probability) for the periodograms of set 2. As expected, periodograms for which the methods assign a probability close to 1 are predominately originated by real planets. Inversely, outputs very close to zero are mostly associated with periodograms containing no planets. 
On the other hand, the value of the x-axis where both histograms cross is much larger for the FAP method.

\subsection{Defining the threshold for planet detection}
\label{lbl:thrsearch}

Once the general performance has been analyzed, we now turn to the topic of finding an adequate threshold for the neural network model. As was already mentioned we considered a log threshold of -1.3, i.e. a 0.05 probability for the FAP. The number of false positives and false negatives produced by the FAP method using this threshold was computed and used to find the range of probability thresholds for the network that gives a smaller number of both false positives and false negatives.

\begin{figure*}
	\begin{subfigure}{.49\textwidth}
		\includegraphics[width=1\linewidth]{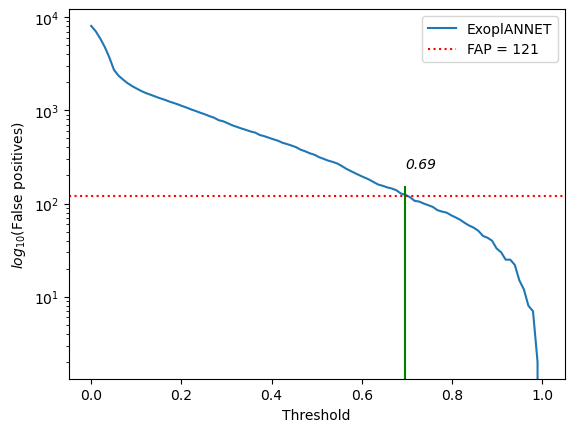}
		\caption{False positives as a function of the threshold.}
	\end{subfigure}
	\begin{subfigure}{.49\textwidth}
		\includegraphics[width=1\linewidth]{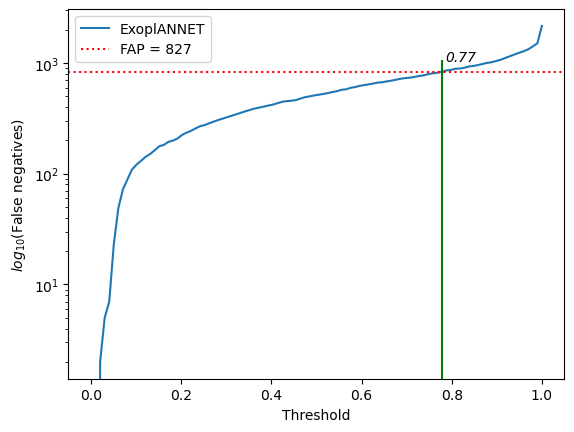}
		\caption{False negatives as a function of the threshold.}
	\end{subfigure}
        \hfill	
        \caption{Lower and upper limit of the threshold where the network has both fewer false positives and false negatives than the traditional method. The dashed horizontal lines show the number of wrong detections with the FAP method, and the vertical segment aids in the identification of the corresponding probability threshold.}
	\label{fig:test}
\end{figure*}

In Figure \ref{fig:test}, the number of false positives and false negatives are shown as a function of the threshold used for the network. The values produced by the FAP method are marked as horizontal dashed lines. This defines a range between 0.69 and 0.77 within which the network obtains better results than the FAP. Any threshold value greater than 0.69 will give fewer false positives, and those less than 0.77 will give fewer false negatives. The network threshold was then set to 0.77, in order to produce the most reliable detection mechanism in terms of false positives, with a similar recall as the traditional method.

\subsection{Performance on test set}
\label{lbl:performanceontest}

The last performance evaluation was conducted on set 3. Since set 2 was used to identify the threshold range where the network performs better than the traditional method, there is a concern that the choice of the threshold may have been influenced by the characteristics of this particular set, leading to overfitting. To address this, we evaluated the network's performance on the yet-unseen set 3.

The \textit{confusion matrices} for both methods are shown in tables \ref{tbl:mc_1} and \ref{tbl:mc_2}. The rows show the true labels of positive and negative cases in the set and the columns show the predicted label.

\begin{table}
    \centering
    \begin{subtable}[c]{.5\textwidth}
	\centering
	\parbox{.5\linewidth}{
		\begin{tabular}{lllll}
			& \multicolumn{4}{c}{Prediction}                                                \\ \cline{3-4}
			& \multicolumn{1}{l|}{}  & \multicolumn{1}{c|}{0}     & \multicolumn{1}{c|}{1}    &  \\ \cline{2-4}
			\multicolumn{1}{l|}{Real Value} & \multicolumn{1}{l|}{0} & \multicolumn{1}{l|}{\textcolor{green}{15309}} & \multicolumn{1}{l|}{\textcolor{red}{277}}  &  \\ \cline{2-4}
			\multicolumn{1}{l|}{}           & \multicolumn{1}{l|}{1} & \multicolumn{1}{l|}{\textcolor{red}{1672}}  & \multicolumn{1}{l|}{\textcolor{green}{2742}} &  \\ \cline{2-4}
		\end{tabular}
		\caption{Confusion matrix FAP}
		\label{tbl:mc_1}
	}
     \end{subtable}
    \hfill
     \begin{subtable}[c]{.5\textwidth}
     	\centering
	\parbox{.5\linewidth}{
		\centering
		\begin{tabular}{lllll}
			& \multicolumn{4}{c}{Prediction}                                                \\ \cline{3-4}
			& \multicolumn{1}{l|}{}  & \multicolumn{1}{c|}{0}     & \multicolumn{1}{c|}{1}    &  \\ \cline{2-4}
			\multicolumn{1}{l|}{Real Value} & \multicolumn{1}{l|}{0} & \multicolumn{1}{l|}{\textcolor{green}{15388}} & \multicolumn{1}{l|}{\textcolor{red}{198}}  &  \\ \cline{2-4}
			\multicolumn{1}{l|}{}           & \multicolumn{1}{l|}{1} & \multicolumn{1}{l|}{\textcolor{red}{1674}}  & \multicolumn{1}{l|}{\textcolor{green}{2740}} &  \\ \cline{2-4}
		\end{tabular}
		\caption{Confusion matrix \texttt{ExoplANNET}.}
		\label{tbl:mc_2}
	}
      \end{subtable}
		\caption{Confusion matrices \texttt{ExoplANNET} and FAP. The rows show the true labels of positive and negative cases in the set and the columns show the predicted label}
\end{table}

The results obtained in this evaluation are consistent with those from set 2. Compared to the traditional method, the number of false positives decreased by 28.5\%, while only two additional false negatives were identified. This is reflected in the comparison in table \ref{tbl:PE_FAP_Red}, where the recall remains almost unchanged but the precision is increased.

\begin{table}[H]
	\centering
	\begin{tabular}{l|c|c|}
		\cline{2-3}
		& FAP   & \texttt{ExoplANNET}   \\ \hline
		\multicolumn{1}{|l|}{Precision}     & 0.908 & 0.932 \\ \hline
		\multicolumn{1}{|l|}{Recall} & 0.621 & 0.620 \\ \hline
	\end{tabular}
	\caption{\texttt{ExoplANNET} and FAP Precision/Recall.}
	\label{tbl:PE_FAP_Red}
\end{table}

\subsection{Execution time}

It was shown in the introduction that, depending on the result, the calculation of the FAP can be done using the Baluev formula or by a Monte Carlo-type algorithm. The analytical formula is not very computationally expensive but assumes white noise. On the other hand, running the Monte Carlo computation on a large data set of radial velocities and computing their corresponding periodograms can take up considerable time, especially when red noise has to be simulated. All this means that the average time required to calculate a FAP value can be considerable.

The case of the network is quite different. A trained network is nothing more than a large matrix and making a prediction is simply taking the input and executing basic operations such as addition and multiplication to return a value. Because of this simple fact, the network should be much faster \footnote{Just to mention another advantage, the network occupies only 1.7kB in memory.}.

To compare the execution times between the methods, we need to consider the \textit{Wall-Time} and the \textit{Proc-Time}. The latter is the time used exclusively to execute a process. It does not take into account things like disk access or other input/output operations, operating system interruptions, time used in other processes, etc. In contrast, \textit{Wall-Time} is the clock time it takes to execute something. It is simply a stopwatch, so if the process is waiting to read data or the system is very busy with several processes, it will add to this time.

On the other hand, if you want to evaluate the time of small programs or pieces of code, it is usually convenient to see only the \textit{proc-time}. The code for generating the periodograms is not parallel (at least not in its current implementation), so these two times should be very similar.

The methods were executed several times, and the resulting running times were averaged. For the FAP method, we perform 100 executions (note that if Baluev's result is good enough, the Monte Carlo simulations are not necessary). For the network method, we generated predictions for $10^6$ periodograms \footnote{The executions were on a desktop computer with a 2.3 GHz micro Intel Core I5 with 8Gb of RAM.}. The results are presented in table \ref{tbl:time_FAP_red} where we report the time taken to analyse a single periodogram, expressed in seconds.

\begin{table}[H]
	\centering
	\begin{tabular}{l|c|c|}
		\cline{2-3}
		& FAP   & \texttt{ExoplANNET}      \\ \hline
		\multicolumn{1}{|l|}{\textit{Wall-Time}} & 425.97 s  & 0.0011 s\\ \hline
		\multicolumn{1}{|l|}{\textit{Proc-Time}} & 427.69 s   & 0.0034 s  \\ \hline
	\end{tabular}
	\caption{Execution times in seconds for the analysis of a single periodogram with \texttt{ExoplANNET} and FAP methods.}
	\label{tbl:time_FAP_red}
\end{table}

The difference is considerable. Focusing only on \textit{proc-time}, the network is 125,791 times faster, that is, a difference of 5 orders of magnitude. The execution time required to calculate the FAP can seem excessively long, but we remind the reader that the Monte Carlo simulations performed include red noise, in an attempt to make the method as reliable as possible.

Note however that the code to calculate the FAP is not optimized. The noise simulations and the computation of the periodograms are severely affecting performance. These parts of the computation could be optimized, and the Monte Carlo procedure could be parallelized. However, the reality is that the advantage that the network brings is too great for any realistic optimization to bring it even close to these results.

Since we are dealing with a relatively small network model and a manageable data size, the use of a GPU was unnecessary. The training process required about 30 minutes using a processor-optimised version of Keras/TensorFlow. However, when experimenting with larger networks and periodograms, GPU boards would likely become advantageous.

\subsection{Virtual Astronomer}
\label{lbl:astro_virtual}

Having compared the performance of \texttt{ExoplANNET} with the traditional method using the FAP, we now seek to compare how both implementations would work in a fully automated process of sequentially finding planets in a time series.

The procedure often follows these lines: when a significant peak is found  (i.e. its $log_{10}(FAP) \leq -1.3$), a sinusoidal signal at that frequency is fit and removed from the data. A new periodogram is computed from the residuals, and the process is repeated until the largest peak in a periodogram is not deemed significant. The pseudocode of this procedure --called \textit{virtual astronomer}-- is shown in Algorithm \ref{alg:virt_astro}. This procedure is often the first step in analysing a radial velocity time series, although there are much more sophisticated techniques to evaluate the significance of a signal in a dataset \citep[e.g.][]{Diaz2016TheHS}. 

\begin{algorithm}
	\caption{virtual astronomer}\label{alg:virt_astro}
	\begin{algorithmic}[1]
		\Procedure{virt\_astro}{$rad\_vel$} 
		\State$pg \gets $gen\_periodogram($rad\_vel$) 
		\State$max\_peak \gets$ get\_max\_peak($pg$) 
		\State$planets=[]$ 
		
		\While{planet($max\_peak, pg$)} 
		\State$planets$.append$(max\_peak)$
		\State$new\_rad\_vel \gets $remove($max\_peak, rad\_vel$)
		\State$pg\gets$ gen\_periodogram$(new\_rad\_vel)$ 
		\State$max\_peak \gets$ get\_max\_peak$(pg)$ 
		\EndWhile
		
		\State \textbf{return} $planets$\
		\EndProcedure
	\end{algorithmic}
\end{algorithm}

So far, all the tests were performed on individual periodograms, which may have contained any number of simulated planets between zero and four, but how the methods behave when used as part of the \textit{virtual astronomer} was not studied. 

We performed an analysis using the threshold values determined in the previous section. Both implementations of the \textit{virtual astronomer}, one using the classical FAP computation and another one using the CNN predictions, were applied to each of the 5000 stars in the third data set. The idea is to test how many planets the methods actually find in a complete, more realistic use case. In particular, we are interested in exploring their behaviour with respect to false positives. 

While it is clear from the above analysis that \texttt{ExoplANNET} performs better in terms of false positives, it is not so clear that this would also be the case when using it with the \textit{virtual astronomer}. On the one hand, the \textit{virtual astronomer} with the FAP method labels more noise peaks as planets, which means that on average the execution will proceed over more iterations than the implementation using \texttt{ExoplANNET}. This may increase the chances that in those additional iterations, the method detects planets that \texttt{ExoplANNET} could not find because it was better at identifying the first dominant non-planetary peak, which stopped the algorithm. On the other hand, this could also mean more false positives are identified with the FAP method.

\begin{figure}
		\centering
		\includegraphics[width=1\linewidth]{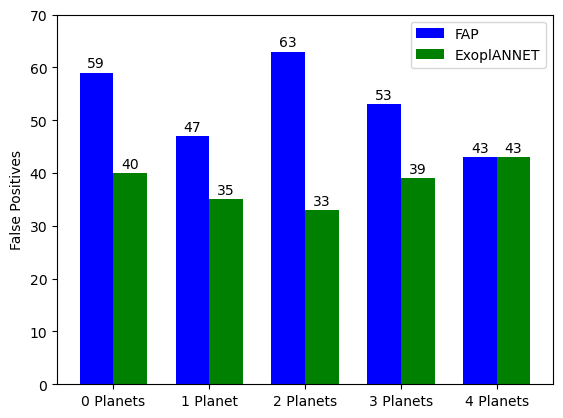}
		\caption{False positives of the \textit{virtual astronomer} with both implementations.}
		\label{fig:fp_virt_astro}
	\end{figure}

To evaluate the performance of the \textit{virtual astronomer} using each detection method, the numbers of false negatives and false positives were compared. In figure \ref{fig:fp_virt_astro} we compare the performances as a function of the actual number of planets in the simulated system. 
The \textit{virtual astronomer} using \texttt{ExoplANNET} yields a lower number of false positives, compared to the implementation with the FAP method. Additionally, the number remains approximately constant over systems with different numbers of planets (\ref{fig:fp_virt_astro}). On the other hand, the results using the FAP exhibit a larger difference for stars with a different number of planets.

\begin{figure*} 
	\begin{subfigure}{.50\textwidth}
 	    \centering
		\includegraphics[width=1\linewidth]{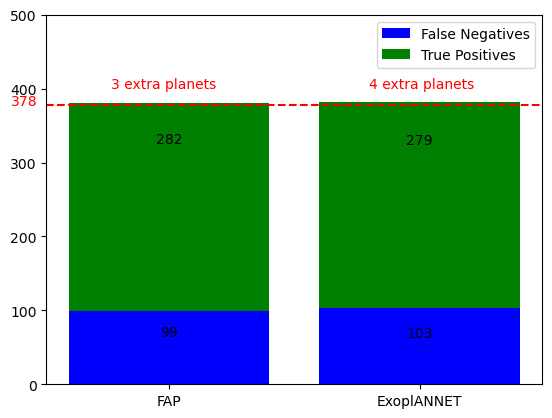} 
		\subcaption*{(a) 1000 stars with 1 planets.} 
		\includegraphics[width=1\linewidth]{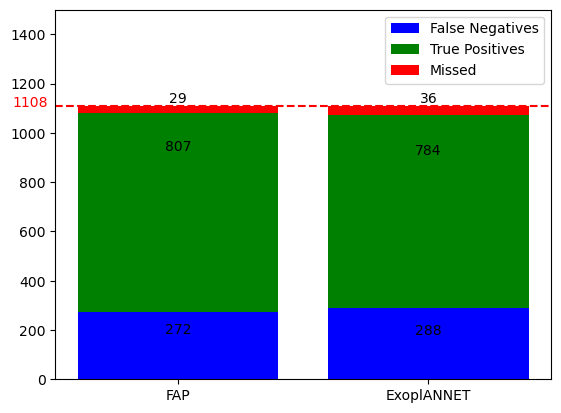}
		\subcaption*{(c) 1000 stars with 3 planets.} 
	\end{subfigure}
	\begin{subfigure}{.50\textwidth}
           \centering	
		\includegraphics[width=1\linewidth]{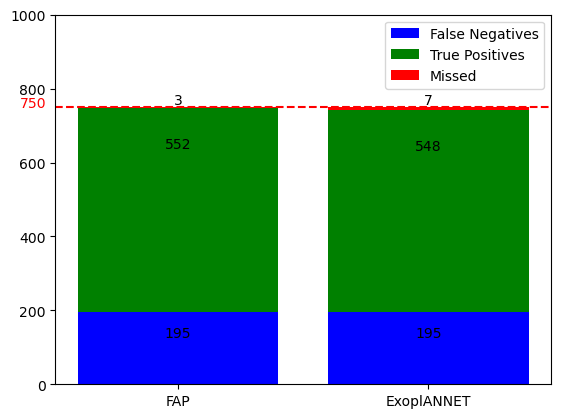} 
		\subcaption*{(b) 1000 stars with 2 planets.} 
		\includegraphics[width=1\linewidth]{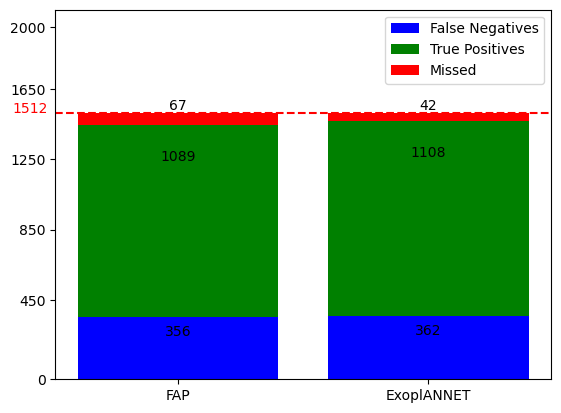}
		\subcaption*{(d) 1000 stars with 4 planets.} 
 	\end{subfigure}
 \caption{Detections of both implementations of the \textit{virtual astronomer}, compared to the detections provided by a hypothetical optimal method with flawless decision making (see text for details). Each panel corresponds to systems with a given number of companions. The numbers of planets detected by the flawless decision-making method are marked by dashed horizontal lines and indicated in red on the vertical axes. The bars present the true positive (i.e. detected planets, green), false negatives (blue) and in red planets that were not detected (i.e. their peak was not evaluated). Note that even using the perfect method, more than half of the planets are missed. Also note that for systems with 1 planet, both FAP and ExoplANNET produce more planetary classifications than the flawless method.} 
	\label{fig7} 
\end{figure*}

To further evaluate both implementations, the numbers of detected planets were compared to the maximum possible number under flawless decision-making, i.e. the maximum number of planets that could be found by the \textit{virtual astronomer} if it had access to the real labels of the peaks. Remember that the fact that a star has \textit{N} planets does not mean that all those planets will be reached by the \textit{virtual astronomer}. If at a given step the peak corresponding to the planet is not the largest peak in the periodogram, the \textit{virtual astronomer} will fail to identify it, and the process will stop. The results are shown in Figure \ref{fig7}.

 As shown by Figure~\ref{fig7}, the \textit{virtual astronomer} using the FAP method permits detecting more planets in the system. However, these extra planets come at the price of a larger number of false positives (Fig.~\ref{fig:fp_virt_astro}). Similarly, the \textit{virtual astronomer} using the FAP method exhibits systematically fewer false negatives, but only by a very small amount. Also, with the exception of the data with four planets, the FAP method also has fewer missed planets, i.e. planets that are not even evaluated by the method. This was expected because of the larger number of iterations brought forth by a larger false positive rate. Overall, the FAP method is more inclined to produce a positive classification, which brings more planets but also more false positives.

Finally, to provide a global view, we computed the accuracy and completeness metrics of each implementation (see Table~\ref{mat:virt_ast_pr}). Despite the difference in the number of periodogram evaluations with each implementation, the precision and completeness values maintain the relationship they had when they were based on single periodogram evaluations. Globally, the completeness is almost the same but the implementation with \texttt{ExoplANNET} increases the precision of the results, significantly reducing the number of false positives.

\begin{table}[H]
	\centering
	\begin{tabular}{l|c|c|}
		\cline{2-3}
		& \begin{tabular}[c]{@{}c@{}}\textit{Virtual Astronomer}\\ FAP\end{tabular} & \begin{tabular}[c]{@{}c@{}}\textit{Virtual Astronomer}\\ ExoplANNET\end{tabular} \\ \hline
		\multicolumn{1}{|l|}{Precision}     & 0.911                                                           & 0.934                                                           \\ \hline
		\multicolumn{1}{|l|}{Recall} & 0.747                                                           & 0.741                                                           \\ \hline
	\end{tabular}
	\caption{Precision and Recall of the \textit{Virtual Astronomer} using the classical FAP method and ExoplANNET to decide on the nature of periodogram peaks.}
	\label{mat:virt_ast_pr}	
\end{table}

\section{Discussion}
\label{sec:Discussion}

\begin{figure*}
	\centering
	\begin{subfigure}{.5\textwidth}
		\centering
		\includegraphics[width=1\linewidth]{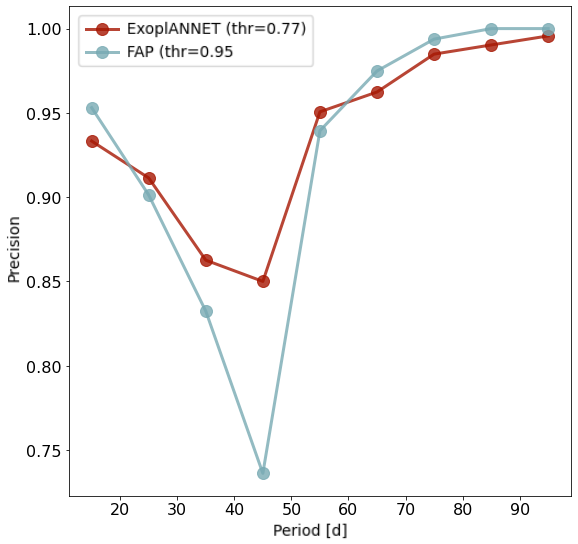}
		\caption{Precision depending on the period.}
	\end{subfigure}%
	\begin{subfigure}{.5\textwidth}
		\centering
		\includegraphics[width=1\linewidth]{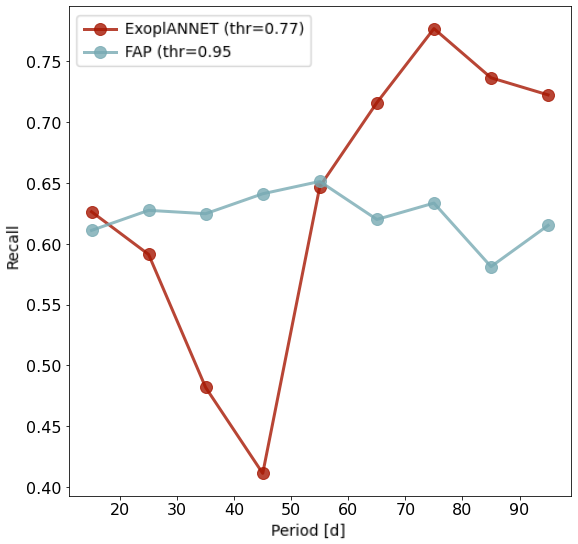}
		\caption{Recall depending on the period.}
		
	\end{subfigure}
	\caption{Precision and recall as a function of the period of the detected planets, using the threshold (thr) 0.77 for the network and 0.95 for the FAP (see Sect.~\ref{lbl:thrsearch}.})
	\label{fig:pres_recall_x_periodo}
\end{figure*}

\begin{figure*}
	\centering
	\begin{subfigure}{.5\textwidth}
		\centering
		\includegraphics[width=1\linewidth]{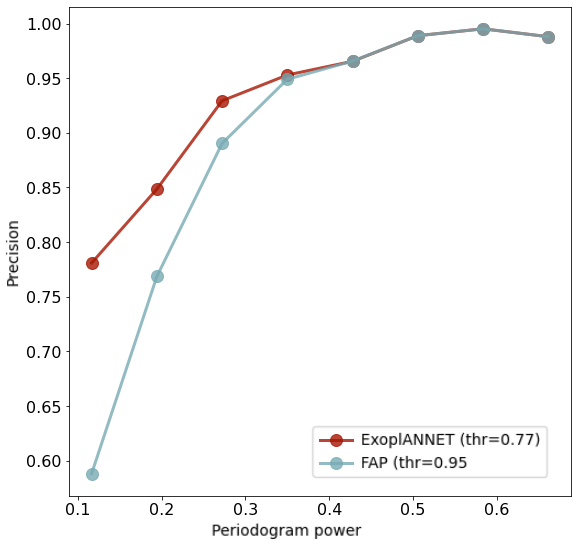}
		\caption{Precision depending on the power.}
	\end{subfigure}
	\begin{subfigure}{.5\textwidth}
		\centering
		\includegraphics[width=1\linewidth]{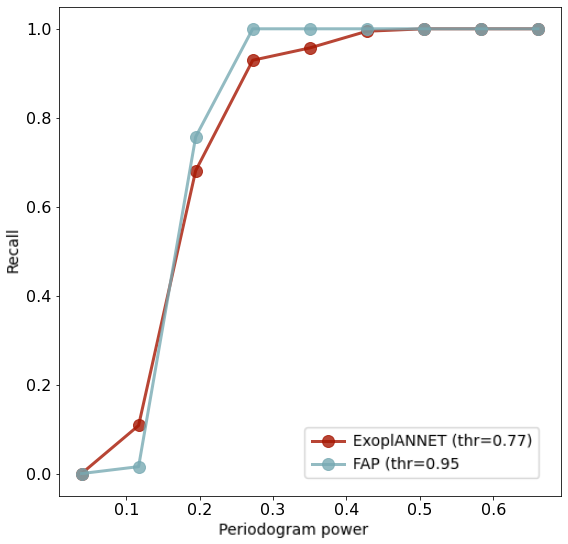}
		\caption{Recall depending on the power}
	\end{subfigure}
	\caption{Precision and recall as a function of the power in the periodogram of the detected planets, using the threshold (thr) 0.77 for the network and 0.95 for the FAP.}
	\label{fig:pres_recall_x_potencia}
\end{figure*}

Having analyzed the general behaviour of the FAP and \texttt{ExoplANNET} as methods for identifying \emph{bona fide} planets in the presence of correlated noise, we now turn to explore the dependence of the performance of the methods with planetary parameters.

In Figure \ref{fig:pres_recall_x_periodo} we plot the precision and recall metrics as a function of the periods of the signals identified as planets, binned every five days. Both methods lose precision in the period range between 30 and 60 days. But while the precision of the FAP method drops below 0.75, the precision of ExoplANNET predictions stays above 0.85. On the other hand, the FAP shows a relatively constant recall across orbital periods, while the network exhibits a clear decrease at around the same period range.

This period range is associated with stellar rotation, and therefore many false positive signals are expected to appear here. Our interpretation is that ExoplANNET learns to be more ``cautious'' when labelling peaks appearing in this range, which incidentally is the same way a professional astronomer usually behaves. As a consequence, the recall is degraded but the precision is kept relatively high. The FAP, on the other hand, does not exhibit any dependence in recall with orbital period and is therefore labelling more noise peaks as planets. As a consequence, the precision is degraded even more than for the network. 

In Figure \ref{fig:pres_recall_x_potencia} precision and recall are plotted as a function of the periodogram peak powers, binned every 10 units of relative power. Although the curves for both methods are  similar, for low-power peaks ExoplANNET is more precise. This leads us to think that the network must be considering characteristics of the periodogram other than the maximum peak power, which allows it to better identify the relatively faint planetary peaks. It is in this range where the network produces less false detection than the FAP method, which explains the results of the previous sections.

\subsection{Robustness against simulation parameters}

The robustness of the results presented in previous sections was tested against changing the number of data points in the test time series and  against the presence of gaps in the time series or a less uniform time sampling. 

To do this, we designed an experiment in which the test time series were modified in different ways:

\begin{enumerate}
    \item A number of points were removed randomly from the time series. This allowed us to test the change in the performance metric against a change in the number of points.

    \item A block of a number of observations was removed from a random position in the time series.

    \item A number of points were removed from the beginning and end of the time series. Half the number of points is removed from each side of the time series.
\end{enumerate}

We experimented with different number of points (20, 40 and 60), and evaluated the model on the modified test time series  without re-training the algorithm. We measured the precision and recall in all cases The results are reported in Table~\ref{tbl:robust}

\begin{table*}
\centering
\caption{Performance of ExoplANNET under different simulated time series sampling and number of points. Values are to be compared with the standard precision (0.932) and recall (0.620) using the same threshold.}
\label{tbl:robust}
\begin{tabular}{c|cc|cc|cc}
\hline\hline
     Method &  \multicolumn{2}{c}{20 points} & \multicolumn{2}{c}{40 points} & \multicolumn{2}{c}{60 points}\\
     & Prec. & Recall &  Prec. & Recall & Prec. & Recall\\
     \hline
     1 & 0.930& 0.608 & 0.927 & 0.610 & 0.889 & 0.648\\
     2 & 0.929& 0.597 & 0.886 & 0.639 & 0.839 & 0.697\\
     3 & 0.929& 0.622 & 0.899 & 0.659 & 0.849 & 0.703\\
     \hline
\end{tabular}
\end{table*}

Overall, we see that the performance of the network is only degraded slightly when the sampling and the number of points of the test set are changed. Across all strategies, the precision decreases as more points are removed. Also, randomly removing points seem to affect precision less than removing them in block, which in turn has a lesser effect than removal from the ends of the time series. This is reasonable, as strategies 1 and 2 mostly retain the time span of the observations, which is a key parameter in determining the shape of the peaks appearing in the periodogram, while data sets modified using strategy 3 will produce periodograms with peaks which are somehow broader and may be hindering correct classification.

On the other hand, the trends of recall are inverted. As the number of points being removed increases, the recall actually improves. This is likely caused by the fact that with the lesser points, a larger fraction of small-mass planets become undetectable --i.e. cease producing the largest peak in the periodogram-- compared to more massive ones. Because the network exhibits a better performance on larger planets, the reduced data set leads to an increase in recall.

\subsection{Recovery of a real signal}
The network was trained on simulated data with particular characteristics of sampling, noise and planetary signals. To test the ability of the network to recover real signals in actual data sets, we employed the HARPS dataset of HD 40307. 

HD 40307 is known to host at least four low-mass planets \citep[e.g.]{mayor2009, tuomi2013, diaz2016a}, one of which has an orbital period of 9.6 days, well within the period range on which the network was trained. We employed the HARPS data set from \citet{diaz2016a} which has 441 measurements over more than 10 years.

\begin{figure*}
     \centering
     \begin{subfigure}[b]{0.48\textwidth}
         \centering
 		    \includegraphics[width=1\linewidth]{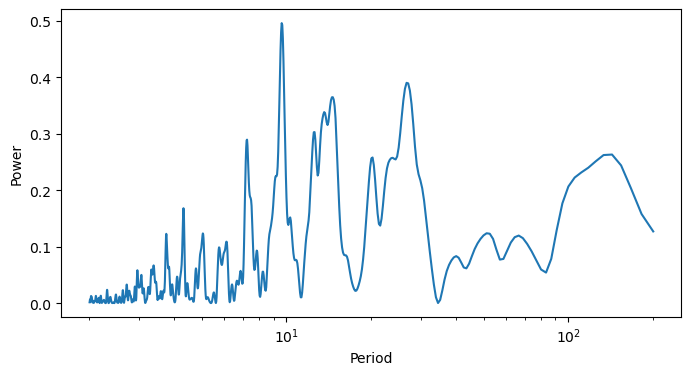}\hfill%
         \caption{HD40307 - Trimmed \\ ExoplaNNET prediction = 1.0}
         \label{fig:_3}
     \end{subfigure}
     \hfill
     \begin{subfigure}[b]{0.48\textwidth}
         \centering
			\includegraphics[width=1\linewidth]{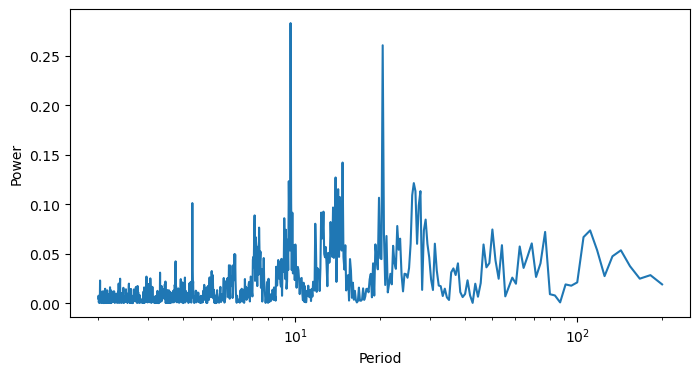}
         \caption{HD40307 - Full \\ ExoplaNNET prediction =0.77}
         \label{fig:_4}
     \end{subfigure}
     \hfill
        \caption{Periodograms of HARPS data of HD40307. \emph{(a)}: trimmed to resemble the sampling of the training set as much as possible; \emph{(b)}: using full dataset.}
        \label{fig:hd40307}
\end{figure*}

We looked for a subset of the data set with characteristics similar to the simulated sets and found a period of 190 days with 220 measurements between Julian dates 53692.7 and 53882.5. We constructed the periodogram for the data in this range of dates and computed a prediction using the network. The predicted probability for the peak at 9.6 days, the largest in the periodogram was very close to 1 (see Fig.~\ref{fig:hd40307}). 

To see how this value varied under a different dataset for the same star, we recomputed the periodogram using the entire data set. As shown in figure \ref{fig:hd40307}, the peaks become much narrower, as expected. The absolute peak power is also decreased in this dataset\footnote{Remember the peak power is a comparison of the $\chi^2$ of a sinusoidal model to a constant one.}. The network predicted probability decreases to around 0.77, likely because of the differing peak width and power.

Unfortunately, the requirement of having datasets with around 200 measurements in a time window of around 200 days limits severely our ability to further test the method on real data. This would require a modification of the CNN architecture, which is outside the scope of this article.

\subsection{Exploring the network caution around activity periods.}

As shown in Sect.~\ref{sec:Discussion}, one of the differences between the procedure using ExoplANNET and the traditional method is that the network does not treat peaks at all frequencies in the same way. Indeed, the network exhibits a decrease in precision around the periodicity associated with activity signals, but much smaller than the one shown by the FAP method. This is compensated by a decrease in recall around these periods. In other words, the network seems to be "more cautious" when judging a peak at the vicinity of the typical rotational period.

\begin{figure}
		\centering
		\includegraphics[width=1\linewidth]{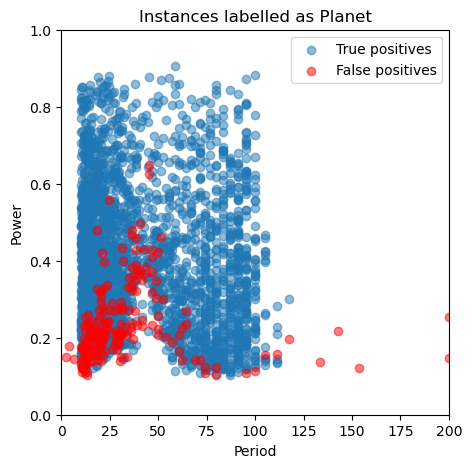}
		\caption{Power and period of largest peaks in periodograms of set 3. The figure shows instances labelled by ExoplANNET as Planet, and blue points indicate instances correctly classified while red points show classification errors.}
		\label{fig:pred_as_planet}
	\end{figure}

To explore this further, we studied the distribution of peak powers and periods for those instances labelled by the network as planets. In Fig.~\ref{fig:pred_as_planet} we present this distribution. A wedge is clearly missing around the periods of the activity signals, where the network requires, on average, a larger power to declare that a signal is a planet. Note, however, that peak power is not the only criterion, as some instances with a much smaller power are classified correctly as planets.

Unfortunately, because of the way the data was constructed, we do not have access to the actual activity period injected for each case. Therefore, we cannot test whether the cautious range decided by the network depends on the actual activity period. Future work would focus on understanding this behaviour in much more detail. Additionally, the possibility of including activity proxies as features for the network may also help in recovering small signals even in the range of periods associated with stellar activity.
\section{Conclusions}

We implemented a Convolutional Neural Network model (named \texttt{ExoplANNET}) to evaluate the presence of planetary signals in radial velocity periodograms. The algorithms was trained on simulated data including red noise and various planetary companions. Its performance was evaluated using previously unseen data and compared with the traditional method based on null  hypothesis significance testing, a.k.a. false alarm probability (FAP). 

Our method outperforms the FAP method on single periodograms, as evaluated by the precision-recall area-under-curve metrics. In other words, a threshold can be chosen, so that it produces a similar level of recall with about 30\% less false positives. In addition, the network is at least five orders of magnitude faster than the traditional FAP method. However, remember that in this work, the noise realizations used to compute the significance of observed peaks are performed using correlated noise, which is certainly slower than methods using only white noise, and could in addition be further optimised by using dedicated methods \citep[e.g.][]{CELERITE, GEORGE}. Finally, \texttt{ExoplANNET} employs barely 1.7kB of computer memory. The trained network is provided in h5 format in the repository \texttt{https://github.com/nicklessagus/ExoplANNET}, together with a sample of labelled periodograms for testing.

The detection method was implemented within an iterative procedure to sequentially explore the presence of planetary signals in a radial velocity data set. This ``virtual astronomer'' evaluates the maximum peak present in a given periodogram and if it decides it belongs to a planet removes a sinusoidal signal at the detected period and proceeds using the model residuals. If a peak is classified as not belonging to a planet, the procedure stops. The FAP method was again used as comparison for the performance of \texttt{ExoplANNET} used as part of the virtual astronomer. Overall, our new method exhibits a better precision metric than the FAP (0.93 vs. 0.91), while keeping a similar recall metric. 

However, using the virtual astronomer with the FAP method more planets are detected, at the expense of a larger number of false positives. It is probably the higher false positive rate that allows the virtual astronomer with the FAP method to evaluate periodograms which follow from a false detection. The same periodogram might be ignored by the virtual astronomer using the \texttt{ExoplANNET} model if the first peak was correctly identified as noise. Indeed, we see that the number of missed planets (i.e. planets that are not even evaluated by the algorithm) is larger under \texttt{ExoplANNET}. 

The behaviour of the detection metrics as a function of  the period of the signal being detected shows a qualitative difference between \texttt{ExoplANNET} and the traditional FAP method. While the FAP retains a relatively constant recall across the entire period range, the neural network seems to have learnt to be more cautious in the period range associated with stellar rotation --and hence activity. As a consequence, while both methods lose precision at this period range, \texttt{ExoplANNET}'s precision remains larger than the one computed with the FAP. This is precisely the reason implementations using machine learning algorithms are useful. As mentioned in the Introduction, the objective is to bypass the modelling of the stellar activity and being able to distinguish between bona fide planets and spurious RV signals nevertheless. In this sense, the comparison with the FAP method is unfair: the traditional method is not even intended to distinguish activity from planets, but merely to claim statistical significance of a signal. Our machine learning method is therefore producing two results at the same time: assessing the significance of a signal and identifying its nature.

The model robustness was studied by removing measurements from the simulations with different criteria. The results show that the network retains much of its performance even when 60 points (30\%) are removed, independently of how they are removed. 

On the other hand, application on real signals shows that, while the network adequately identifies a known planet, its performance would likely be degraded if employed on long time series typical of RV surveys. This is likely due to the much narrower peaks associated with long time series compared to the simulated data sets used for training. Future work will address this limitation, which would require significant changes both in the network architecture and in the data set simulation procedure.

Future work shall focus on evaluating the performance on a larger sample of real-time series. This will be done both by performing simulations with realistic time sampling, and on actual observed data. Taking into account that different sampling and number of points are crucial parameters in the simulations. In addition, neural networks can be provided with rich context to the classification task in the form of additional data. In our case, \texttt{ExoplANNET} takes the position and power of the largest peak in the periodogram. Additional information can come in the form of the stellar rotational period and spectral type, for example. Implementing and exploring these possibilities will constitute avenues for future advancement.
%
%
\begin{acknowledgements}
Part of this work was carried out at the Geneva Observatory and financed by the Seed Money project Grant 2018 ``Towards the Detection of Earth Analogues (TDEA)'' of the STATE SECRETARIAT FOR EDUCATION, RESEARCH AND INNOVATION, Switzerland.
\end{acknowledgements}
\bibliographystyle{aa} 


\begin{appendix}
\section{Descarted approaches} \label{apendix_a} 

During the design phase of ExopANNet, several combinations of convolutional and pooling layers were explored before reaching the final architecture. Overall, the results obtained with different numbers of layers and filters were very similar. It quickly became apparent that the difficulty of the problem did not pose a significant challenge for this type of network, and therefore the most simple combination that produced good results was chosen. 

However, the main issue was determining the appropriate way to model the problem. Many options were considered before arriving at the proposed solution of using CNNs on periodograms:

\begin{itemize}
    \item Initially, the approach involved using the radial velocity time series and treating the problem as a classification task with each number of planets representing a separate class. Convolutional and LSTM networks were experimented with. However, it became clear very quickly that this classification approach did not fit the problem well, as the classes were not mutually exclusive. Consequently, the tests did not yield any substantial improvements over random classification.
    
    \item After discarding the classification approach, the focus shifted towards utilizing periodograms and treating the problem as a linear regression task. In this approach, the network was designed to receive a periodogram as input and generate a real number between 0 and 4, representing the estimated number of planets present. Although the network demonstrated the ability to differentiate between periodograms with and without planets, the results were somewhat imprecise. Further analysis revealed an interesting pattern: the network appeared to be performing a form of ``sum of powers'' calculation on the peaks. Periodograms with higher peak values tended to yield numbers closer to 4, while lower peak values resulted in smaller output values, regardless of the actual number of peaks present in the periodogram.
       
    \item Before the proposed solution outlined in this work, an alternative test was conducted. The initial idea was to input the periodogram directly into the network, which would then identify the peaks associated with planets and output the position of those peaks. While this approach was ambitious, it posed significant challenges in terms of validation and practical implementation. One of the main difficulties arose when the network predicted a position between two peaks, making it impossible to determine which peak should be chosen. Consequently, this approach was deemed infeasible in practice.

\end{itemize}

Although these tests were inconclusive, they served to better understand the problem and the capabilities of the network and led us to find the proposed solution.

\section{Convolutional Neural Networks} \label{apendix} 

In this Appendix, we make a very brief summary of general concepts and usage of convolutional networks. Notions such as layer or backpropagation, and the generalities of artificial neurons (activation function, synaptic weight, bias, etc.) are assumed to be known. Those wishing to investigate further into theoretical and implementation concepts can refer to the bibliography (\cite{bishop2007, 10.5555/104000, Goodfellow-et-al-2016}).

Deep fully-connected networks have achieved very good practical results used in a wide variety of problems. However, for the treatment of images, they have a disadvantage, they do not take into account the spatial structure of the image. As a consequence, they can become extremely inefficient. On the other hand, convolutional neural networks (CNN) were designed specifically to tackle tasks of computer vision. These networks use convolutions instead of classical matrix operations and were created to work with images, but are also useful for one-dimensional data where the structure of the data is relevant.

Broadly speaking, CNNs perform the following steps when processing an image:
\begin{itemize}
	\item{\textit{Feature extraction}}: Each neuron takes its synaptic inputs from a local receptive field, i.e. a small region in the previous layer, and combines them to extract local features.
	\item{\textit{Feature mapping}}: Each layer of the network is made up of many feature maps, each in the form of a plane within which individual neurons are forced to share the same set of synaptic weights. This second form of structural restriction has the following beneficial effects:
	\begin{itemize}
		\item Translational invariance, achieved by performing the convolution operation on a feature map with a small kernel, followed by applying the activation function. A kernel is a small matrix of weights with which the image is convoluted. 
		\item Reduced number of parameters, achieved through the use of shared weights.
	\end{itemize}
	\item{\textit{Subsampling}}: Convolutional layers are often followed by a layer that performs local averaging and subsampling, thus reducing the resolution of the feature map. This operation has the effect of reducing the sensitivity of the feature map output.
\end{itemize}

In this way, a deep network is obtained that in its first layers detects lines or curves; but, as it progresses, it detects more complex shapes such as a face or a silhouette \citep{zeilerfergus2014}.

\begin{figure}[H]
	\resizebox{\hsize}{!}{\includegraphics[width=0.8\linewidth]{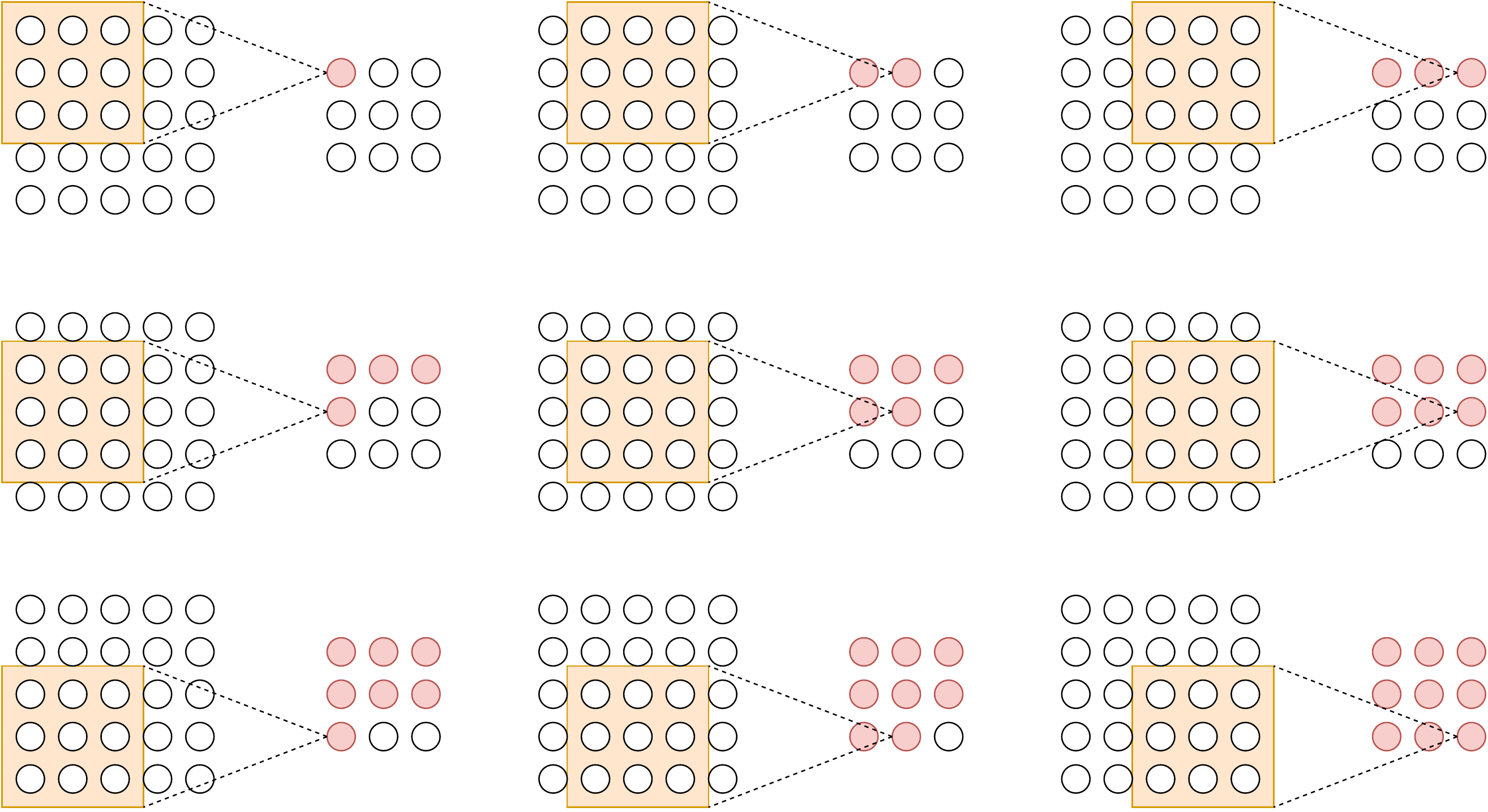}}
	\caption{Generation of a feature map using one kernel.}
	\label{fig:kernel-conv}
\end{figure}

To make a feature map, a kernel is applied to an image, this kernel on the image is the \textit{receptive field} that a neuron ``sees''. In Figure \ref{fig:kernel-conv}, there is an example of how a characteristics map is generated. In the example, a 3x3 kernel is applied on a 5x5 image \footnote{For simplicity, these examples are presented on matrices in two dimensions, assuming black and white images. If they were colour images it would be necessary to add a dimension to the input and the kernel. So we would have three filters, these three filters are added together with the \textit{bias} and they will form an output as if it were a single channel.}.

The size of the kernel is one of the parameters chosen when creating the network. They are usually small to extract local characteristics. To find many of these features, many kernels are used. They all have the same dimension, but the weights they learn are different (they look for different things). This set of kernels is what is called a filter.

Next, we focus on how downsampling layers work and why they are needed, using the classic  MNIST dataset \citep{deng2012} as an example. This dataset consists of tens of thousands of handwritten digit images, which are in black and white and with a resolution of 28x28 pixels. Each image represents a digit between zero and nine.

\begin{figure}[H]
	\resizebox{\hsize}{!}{\includegraphics[width=0.9\linewidth]{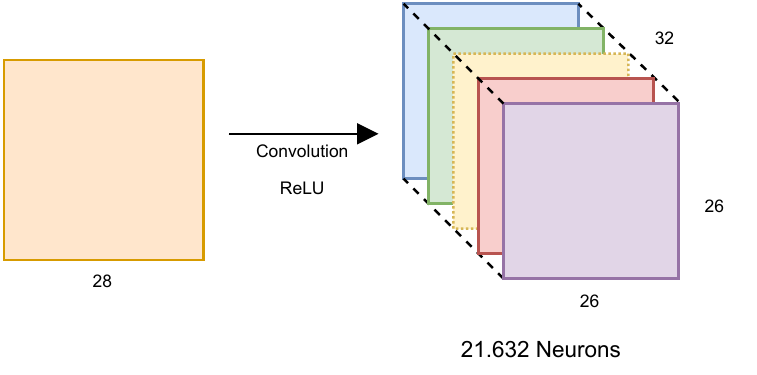}}
	\caption{First hidden layer.}
	\label{fig:feature-map}
\end{figure}

In Figure \ref{fig:feature-map} you can see how applying a filter of 32 kernels of 3x3 gives 32 activation maps of 26x26 each \footnote{The output dimension $o$ to apply a kernel of dimensions $k$ to an image of size $w$ can be calculated with this formula: $o = w-k + 1$. Concepts such as padding or stride modify this formula, but they are unnecessary for the example.}, so the number of neurons in this first hidden layer is 21,632.

The size of the information after the first layer has increased considerably. Consider this is a very small image, and it is only the first layer. Adding more layers can make this number grow very fast. Eventually, this hinders training a network on large high-definition images, since the network will lose the ability to abstract. This is where the downsampling layer comes in. 

\begin{figure}[H]
	\centering
	\includegraphics[width=0.5\linewidth]{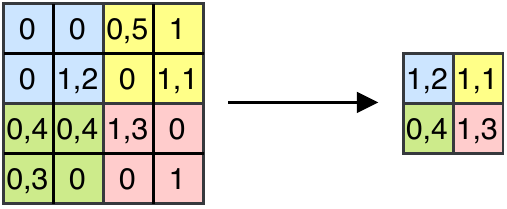}
	\caption{\textit{Maxpooling}.}
	\label{fig:maxpooling}
\end{figure}

The most used implementation is maxpooling. In this mechanism, we seek to reduce the parameters by dividing the input matrix into parts and keeping the maximum value of each one. It seeks to preserve the most important characteristics and reduce the dimensions of the data. In Figure \ref{fig:maxpooling} there is an example of its application.

\begin{figure}[H]
	\resizebox{\hsize}{!}{\includegraphics[width=1\linewidth]{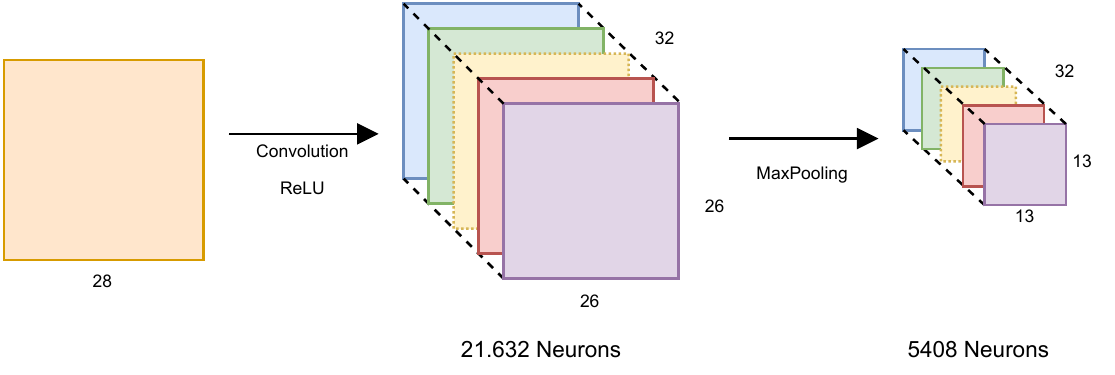}}
	\caption{Use of maxpooling for dimensionality reduction.}
	\label{fig:full-conv}
\end{figure}

Figure \ref{fig:full-conv} shows the result of using this layer over the output of Figure \ref{fig:feature-map}.

\begin{figure}[H]
	\resizebox{\hsize}{!}{\includegraphics[width=0.9\linewidth]{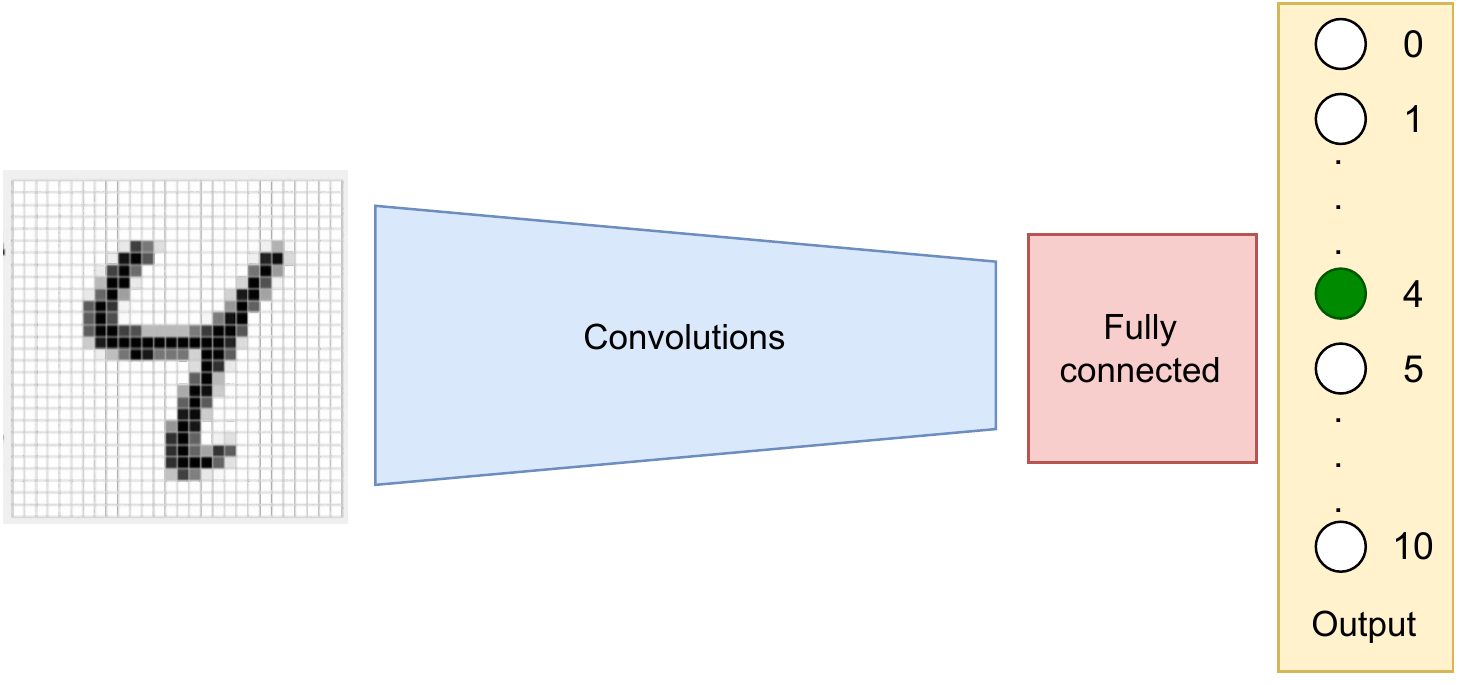}}
	\caption{Full network.}
	\label{fig:conv-complete}
\end{figure}

As expected, the number of layers required to reach good performance, and the manner in which they are interleaved, depend on the problem to be solved, and there is no general rule. But in general, the convolutional network extracts the most relevant features from the image and then connects them to one or more fully connected layers and an output layer used for classification. This layer depends on the problem. In this particular example, it has a neuron for each category --i.e. ten neurons-- and uses a softmax activation function. This activation function normalises the outputs of the output layer so that each value can be interpreted as the probability of belonging to a given class. The fully-connected part, together with the output layer uses the information extracted in the convolutional part, and ``decide'' to which category the entry belongs. Figure \ref{fig:conv-complete} shows a general diagram of the network in this example.

\end{appendix}
\end{document}